\documentclass[aip,jcp,reprint,twocolumn]{revtex4-1}

\draft 

\usepackage{graphicx}
\usepackage{color}
\usepackage{hyperref}
\usepackage{amssymb}
\usepackage{setspace}

\begin{document}

\title{Electronic spectra of linear HC$_5$H and cumulene carbene H$_2$C$_5$}

\author{Mathias Steglich}
\email{m.steglich@web.de}
\affiliation{Department of Chemistry, University of Basel, Klingelbergstrasse 80, CH-4056 Basel, Switzerland}
\author{Jan Fulara}
\affiliation{Department of Chemistry, University of Basel, Klingelbergstrasse 80, CH-4056 Basel, Switzerland}
\affiliation{Institute of Physics, Polish Academy of Sciences, Al. Lotnik\'ow 32/46, PL-02-668 Warsaw, Poland}
\author{Surajit Maity}
\author{Adam Nagy}
\author{John P. Maier}
\email{j.p.maier@unibas.ch}
\affiliation{Department of Chemistry, University of Basel, Klingelbergstrasse 80, CH-4056 Basel, Switzerland}

\begin{abstract}
The $1 ^3\Sigma_u^- \leftarrow X^3\Sigma_g^-$ transition of linear HC$_5$H (\textbf{A}) has been observed in a neon matrix and gas phase. The assignment is based on mass-selective experiments, extrapolation of previous results of the longer HC$_{2n+1}$H homologues, and density functional and multi-state CASPT2 theoretical methods. Another band system starting at 303\,nm in neon is assigned as the $1 ^1 \textnormal{A}_1 \leftarrow \widetilde{X} ^1 \textnormal{A}_1$ transition of the cumulene carbene pentatetraenylidene H$_2$C$_5$ (\textbf{B}).
(\textit{published in J. Chem. Phys. 142 (2015) 244311})
\end{abstract}

\maketitle

\section{Introduction}
Unsaturated hydrocarbon chains and their cyano-substituted analogues take part in astrochemical processes as many of them have been detected in different astrophysical environments, such as dense and diffuse clouds or circumstellar shells. On earth, they play major roles as intermediates in combustion and plasma chemistry as well as in other soot-forming processes. The hydrogen-capped HC$_n$H linear chains are important members of these compounds. They possess no permanent dipole moment and can therefore only be detected via vibrational or electronic spectroscopic methods. For instance, the presence of the even-numbered polyacetylenes HC$_{2n}$H ($n=1-3$) in stellar outflows and planetary atmospheres was inferred from observations in the infrared.\cite{ridgway76,kunde81,kim85,cernicharo01} The odd-numbered counterparts are more difficult to synthesize and study in the laboratory because of their open-shell character. They have to be embedded in collision-free environments, i.e., low-temperature matrices or supersonic jets. Their presence in space, albeit yet unproven, is suggested from astronomical detections of linear analogues, such as C$_{n}$ ($n=3,5$) and HC$_{n}$ ($n=3,5,7$).\cite{hinkle88,maier01,bernath89,thaddeus85a,cernicharo86,guelin97}

Within this context, studies of the first allowed electronic transition $1 ^3\Sigma_u^- \leftarrow X^3\Sigma_g^-$ of the HC$_{2n+1}$H series, situated in the visible or near-infrared, are of relevance. Respective absorption spectra for chains containing up to 19 carbon atoms have been measured in neon matrices ($n=2-7,9$) and in the gas phase by resonant two-color two-photon ionization (R2C2PI; $n=3-6,9$) and cavity ring down spectroscopy (CRDS; $n=3-6$).\cite{fulara95,ball99,ball00,ding03} Attempts to measure the $1 ^3\Sigma_u^- \leftarrow X^3\Sigma_g^-$ transition for the smallest members HC$_3$H and HC$_5$H in the gas phase failed so far. Whereas the two C$_3$H$_2$ isomers with C$_{2v}$ symmetry, which are more stable, were studied\cite{madden89,vrtilek90,hodges00,achkasova06,maier11} arguments were given for the non-observation of the electronic transition of HC$_5$H, including low oscillator strength and unfavorable production conditions.\cite{ball00}

\begin{figure}\begin{center}
 \includegraphics[scale=0.41]{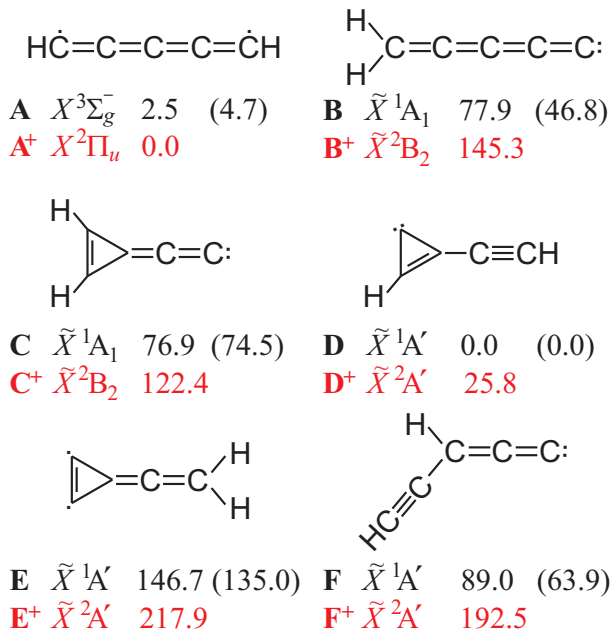}
 \caption{Zero-point corrected ground state energies (kJ/mol) of neutral isomers of C$_5$H$_2$ (black) and their cations (red) at the M06-2X/cc-pVTZ level of theory. Energies calculated at the MS(4)-CASPT2/cc-pVTZ level are given in parentheses.} \label{fig_structures}
\end{center}\end{figure}

Several low-energy isomers of C$_5$H$_2$ were investigated by quantum chemistry and microwave absorption spectroscopy.\cite{seburg97,blanksby98,mavrandonakis02,mccarthy97,travers97,gottlieb98} Coupled-cluster calculations suggest pentadiynylidene (linear HC$_5$H triplet; \textbf{A}) to be the most stable structure, albeit singlet ethynylcyclopropenylidene (c-HC$_3$--C$_2$H; \textbf{D}) is close in energy (8 kJ/mol) and three others (\textbf{B, C, F}; see Fig. \ref{fig_structures}) have been found within 90 kJ/mol.\cite{seburg97} Rotational spectra were recorded for the four lowest singlet isomers possessing a permanent dipole moment, i. e., structures \textbf{B, C, D, F},\cite{mccarthy97,travers97,gottlieb98} whereas the HC$_5$H triplet was studied in solid matrices only.\cite{fulara95,bowling06} The first matrix investigation was realized by co-authors of this paper.\cite{fulara95} Mass-selected C$_5$H$_2$ cations produced in a diacetylene discharge were deposited in cryogenic neon; the linear triplet was subsequently generated by photo-induced neutralization and the electronic absorption spectrum was recorded. In a later study by Bowling et al.,\cite{bowling06} HC$_5$H was created by photolysis of diazo-pentadiyne in a nitrogen matrix; its structure was characterized by IR, EPR, and UV-vis spectroscopy. Inconsistencies between both data sets require a re-investigation of the $1 ^3\Sigma_u^- \leftarrow X^3\Sigma_g^-$ transition.

This article reports the $1 ^3\Sigma_u^- \leftarrow X^3\Sigma_g^-$ spectrum of mass-selected HC$_5$H (\textbf{A}) deposited in solid neon with lower kinetic energy and higher mass-resolution than before. A second absorption system commencing at 303 nm is observed and assigned as the $1 ^1 \textnormal{A}_1 \leftarrow \widetilde{X} ^1 \textnormal{A}_1$ transition of pentatetraenylidene H$_2$C$_5$ (\textbf{B}). Due to band shifts and broadening, electronic spectra of matrix-isolated molecules can not usually be compared with astronomical data for identification purposes. The gas phase detection and R2C2PI characterization of \textbf{A} is reported here for the first time. Vibrational frequencies and rotational constants are obtained.

\section{Methods}
\subsection{Experimental}
Two types of experiments, in neon matrix and gas phase, were carried out. The apparatus used for the former combines mass spectrometry with matrix isolation. It is described elsewhere.\cite{freivogel94} The $m/z = 62$ ions were produced in a hot cathode discharge source using different precursors seeded in helium, such as methyldiacetylene, a mixture of propyne and acetylene, or propyne and diacetylene. Ions extracted from the source were guided to a quadrupole mass filter. Mass-selected C$_5$H$_2^+$ was co-deposited with neon forming a 150 $\mu$m thick matrix containing cations as well as neutrals. The largest C$_5$H$_2^+$ current of about 10 nA was obtained from the methyldiacetylene precursor. Absorption spectra were measured by passing broad band light from a halogen or a high-pressure xenon lamp through the matrix, dispersing the light by a 0.3\,m spectrograph, and recording the signal with a CCD camera. Absorptions from neutrals and cations could be discriminated by further irradiation with UV photons ($\lambda > 260$ nm) leading to enhanced C$_5$H$_2$ formation.

Gas phase spectroscopy was realized by a resonant two-color two-photon ionization scheme (R2C2PI) on a molecular beam of C$_5$H$_2$ and other hydrocarbons. A gas mixture of 0.1\,\% propyne (C$_3$H$_4$) and 0.1\,\% diacetylene (C$_4$H$_2$) in helium discharged in the expansion of a pulsed valve (8 bar backing pressure) provided the target species. Longer homologues of the HC$_{2n+1}$H series were also observed, their low-resolution (0.1\,nm) spectra being identical to those published previously.\cite{ding03} A collimated beam was produced by a skimmer placed 50\,mm downstream. The ions generated by the R2C2PI process were monitored as a function of scanning laser wavelength with a linear time-of-flight mass spectrometer. Spectral scans were undertaken by counter-propagating the radiation of the laser into the molecular beam. An optical parametric laser (5-10\,ns; 20\,Hz; 0.1\,nm bandwidth) provided photons for broad range scans between 410 and 710\,nm. Measured spectra were corrected for wavelength-dependent laser power variations. Higher-resolution scans around 430\,nm were conducted with a dye laser ($\approx 10$\,ns; 10\,Hz; 0.001\,nm bandwidth). The accuracy of the calibration using a wavemeter is $\pm 0.05$\,cm$^{-1}$. An ArF excimer laser emitting 193\,nm photons ($\approx 10$\,ns) perpendicular to the beam ionized the molecules.

\subsection{Computational}
To assign the C$_5$H$_2$ UV features seen in a neon matrix, the six lowest energy isomers depicted in Fig. \ref{fig_structures} were selected for a computational study, whereby structures \textbf{A, B, C, D, F} have been investigated by coupled cluster theory before.\cite{seburg97} As starting point for high-level calculations using a multi-configurational second order perturbation theory (CASPT2), equilibrium geometries were first optimized at DFT level (density functional theory) with the M06-2X functional\cite{zhao08} and the cc-pVTZ basis \cite{dunning89,woon93} set of the Gaussian09 software.\cite{frisch13} The thus obtained structures were then refined with the CASPT2 method of the MOLCAS program\cite{aquilante10} along with the same basis set. The reference wavefunction was optimized for the average electronic energy of four electronic states of a given symmetry (multistate option MS(4)). Ten electrons distributed over ten orbitals (10/10) formed the active space. The obtained ground state energies of neutrals and cations are given in Fig. \ref{fig_structures}. The relative stabilities are comparable to the coupled cluster approach,\cite{seburg97} except for a reversed order of the almost isoenergetic HC$_5$H triplet and c-HC$_3$--C$_2$H singlet. Vertical excitation energies of isomers \textbf{A--F} were calculated at the MS(4)-CASPT2 (10/12) level. Adiabatic excitation energies have been computed for selected electronic states using MS(4)-CASPT2 (10/10). In addition, electronic transitions were predicted with the symmetry adapted cluster/configuration interaction (SAC-CI) method\cite{nakatsuji78,nakatsuji79} and the cc-pVTZ basis set of Gaussian09. Vibrational progressions of chosen C$_5$H$_2$ transitions and the rotational profile of linear HC$_5$H were simulated using the PGOPHER program.\cite{western10} For this, geometries and vibrations in ground and excited states were predicted with the B3LYP functional\cite{becke88,lee88} and cc-pVTZ basis set using Gaussian09's DFT and time-dependent DFT methods.

\section{Results and discussion}
\subsection{Matrix spectrum of HC$_5$H}
\begin{figure}\begin{center}
 \includegraphics[scale=0.4]{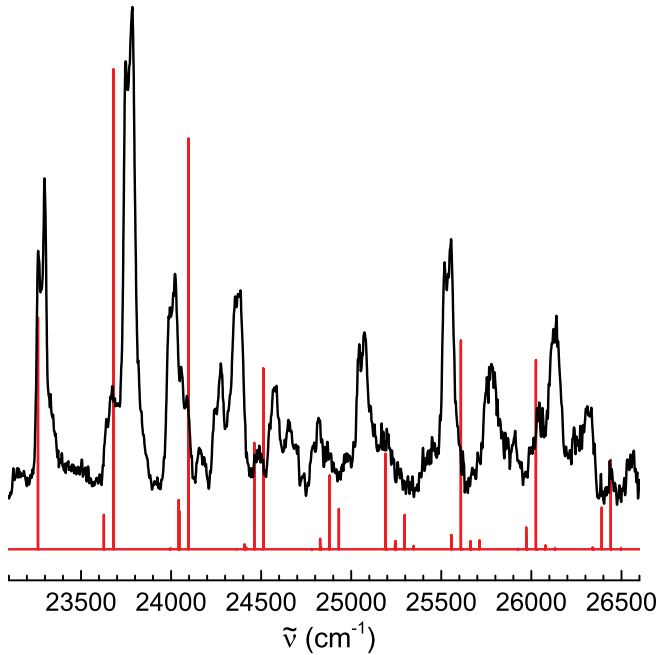}
 \caption{$1 ^3\Sigma_u^- \leftarrow X^3\Sigma_g^-$ spectrum of HC$_5$H in a 6\,K neon matrix compared to a Franck-Condon simulation at the B3LYP/cc-pVTZ level (T$_{vib}=0$ K).} \label{fig_HC5H_MIS}
\end{center}\end{figure}

Deposition of $m/z = 62$ ions in a 6\,K neon matrix leads to absorptions identical with those reported for linear HC$_5$H$^+$.\cite{fulara10} Weak features with onsets at about 430 and 303\,nm gained intensity during UV irradiation. They are therefore associated with neutral species. The 430\,nm system is displayed in Fig. \ref{fig_HC5H_MIS}. It differs from the HC$_5$H neon matrix spectrum published earlier.\cite{fulara95} Back then, absorptions in the 400 -- 435\,nm range were assigned to the $1 ^3\Sigma_u^- \leftarrow X^3\Sigma_g^-$ transition of the linear triplet based on mass-selection and the approximate wavelength of the origin band deduced from the positions of the longer HC$_{2n+1}$H homologues. The most intense bands observed here were also present in the old spectrum, but a stronger absorption system starting at 434 nm assigned recently\cite{fulara15} to H$_2$C$_5$H overlapped the $1 ^3\Sigma_u^- \leftarrow X^3\Sigma_g^-$ transition. This radical was absent in the present investigation because of a better mass resolution and a lower kinetic energy of deposited ions. The correct origin band is now observed at 429.9\,nm (instead of 434\,nm). The difference could be the reason why Ball et al.\cite{ball00} were unable to observe HC$_5$H in their high-resolution study.

The vertical excitation energy of \textbf{A} calculated at the MS(4)-CASPT2(10/12)/cc-pVTZ level is 3.28\,eV ($f=0.005$), quite far away from the measured position of the origin band at 2.88\,eV. A considerable deformation of the chain structure is obtained upon geometry optimization of the excited $1 ^3\Sigma_u^-$ state using MS(4)-CASPT2(10/10). The C--C--H angle is subject to the largest distortion and deviates by about 40$^\circ$ from linearity. The adiabatic excitation energy is 2.85\,eV, which is much closer to the origin band position and comparable to results from a multi-reference single and double excitation configuration interaction method locating the $1 ^3\Sigma_u^- \leftarrow X^3\Sigma_g^-$ transition at 2.76\,eV ($f=0.007$).\cite{mavrandonakis02} No further transition with appreciable oscillator strength is predicted below 6\,eV, but another (dark) $^3\Sigma_u^+$ state is expected in the vicinity of $1 ^3\Sigma_u^-$. The energetic order of these two states is unknown. 

Several vibronic bands in the matrix spectrum show site splitting into doublets of 30 to 40\,cm$^{-1}$ energy difference. The band positions measured at the maximum absorption and assignments based on DFT and CASPT2 calculations are summarized in Table \ref{tab_bands}. Some old assignments\cite{fulara95} have been corrected in light of the new results. A Franck-Condon simulation of the $1 ^3\Sigma_u^- \leftarrow X^3\Sigma_g^-$ transition at the B3LYP/cc-pVTZ level is compared with the matrix absorption spectrum in Fig. \ref{fig_HC5H_MIS}. The computed intensity pattern agrees reasonably well with the observed spectrum, the band positions differ to a larger extent, though. At this level of theory, the fully relaxed excited state structure exhibits C$_2$ symmetry with a C--C--H angle of about 153$^\circ$ on both ends. The carbon frame is slightly wavy; the average C--C distance has increased by about 1.2\,\% compared to the ground state. This implies that the Franck-Condon active vibrations are C--H bending, C--C chain bending, and C--C stretching modes. The strongest band is predicted to be caused by the single excitation of a C--H bending vibration of A symmetry calculated at 417\,cm$^{-1}$. Due to symmetry difference this vibration cannot directly be mapped to a ground state one. Strong Duschinksy mixing is active; the 417\,cm$^{-1}$ mode projects on both C--H bendings $\nu_4(\pi_g)$, $\nu_9(\pi_u)$ of the ground state. For reasons of simplification, the observed bands are labeled within the framework of the ground state, i.e., the strong band at 23747\,cm$^{-1}$ is assigned as $4_0^1$. The vibrational energy of 486\,cm$^{-1}$ is somewhat higher than the calculated value. Higher-order assignments (two or more quanta) are rather tentative; the potential for the C--H bending seems to be highly anharmonic. The C--C chain bending and stretching modes $\nu_5(\pi_g)$ and $\nu_2(\sigma_g)$ appear at 380 and 1770\,cm$^{-1}$ in the excited electronic state in a neon matrix. The computed values in $1 ^3\Sigma_u^-$ are 367 and 1929\,cm$^{-1}$, respectively. The $\nu_2(\sigma_g)$ mode is better reproduced with 1825\,cm$^{-1}$ at the MS(4)-CASPT2(10/10) level.\footnote{The CASPT2 method implemented in MOLCAS allows the computation of only totally symmetric vibrations.} The $3_0^1$ assignment is also tentative due to low Franck-Condon intensity and potential overlap with  the $4_0^15_0^1$ band. The C$-$C stretching mode $\nu_3(\sigma_g)$ is calculated at 767\,cm$^{-1}$ (CASPT2: 755\,cm$^{-1}$), and possibly measured at 801\,cm$^{-1}$.

\begin{table}[h!]
 \caption{Observed absorption bands (cm$^{-1}$) in the $1 ^3\Sigma_u^- \leftarrow X^3\Sigma_g^-$ electronic spectrum of HC$_5$H.}
\begin{tabular}{|ccccc|} \hline
\multicolumn{2}{|l}{Ne matrix}				& \multicolumn{2}{l}{Gas phase}					& Assignment			\\
$\widetilde{\nu}$ 	& $\Delta\widetilde{\nu}$	&	$\widetilde{\nu}$ 	& $\Delta\widetilde{\nu}$	& 						\\ \hline \hline
				 	& 							&	23170			 	& 	-91						& 	$4_1^1$$5_1^1$	\\
				 	& 							&	23202			 	& 	-59						& 	$5_1^1$ or $4_1^1$	\\
				 	& 							&	23234			 	& 	-27						& 	$4_1^1$ or $5_1^1$	\\
23261			 	& 	0						&	23261		 		& 	0						& 	$0_0^0$			\\
23299			 	& 	38						&				 		& 							& 	$0_0^0$			\\
				 	& 							&	23474			 	& 	213						& 	?					\\
23641			 	& 	380						&	23624			 	& 	363						& 	$5_0^1$			\\
23674			 	& 	413						&					 	& 							& 	$5_0^1$			\\
23747			 	& 	486						&	23742			 	& 	481						& 	$4_0^1$			\\
23787			 	& 	526						&					 	& 							& 	$4_0^1$			\\
				 	& 							&	23958			 	& 	697						& 		?				\\
23998			 	& 	737						&	23975			 	& 	714						& 	$5_0^2$			\\
24027			 	& 	766						&					 	& 							& 	$5_0^2$			\\
24062			 	& 	801						&	24050			 	& 	789						& 	$3_0^1$			\\
24091			 	& 	830						&					 	& 							& 	$3_0^1$			\\
24160			 	& 	899						&	24149			 	& 	888						&   $4_0^15_0^1$		\\
24184			 	& 	923						&					 	& 							&   $4_0^15_0^1$		\\
24242			 	& 	981						&	24228			 	& 	967						& 	$4_0^2$			\\
24278			 	& 	1017					&					 	& 							& 	$4_0^2$			\\
				 	& 							&	24272			 	& 	1011					& 	?	\\
				 	& 							&	24295			 	& 	1034					& 	?	\\
24355			 	& 	1094					&	24334			 	& 	1073					& 	$5_0^3$			\\
24390			 	& 	1129					&	24349			 	& 	1088					& 	$5_0^3$			\\
24576			 	& 	1315					&					 	& 							& 	$3_0^1$$4_0^1$	\\
24655			 	& 	1394					&					 	& 							& 	$4_0^3$			\\
24820			 	& 	1559					&					 	& 							& 	$3_0^2$			\\
25031			 	& 	1770					&					 	& 							& 	$2_0^1$			\\
25075			 	& 	1814					&					 	& 							& 	$2_0^1$			\\
25435			 	& 	2174					&					 	& 							& 	$2_0^15_0^1$			\\
25517			 	& 	2256					&					 	& 							& 	$2_0^14_0^1$	\\
25556			 	& 	2295					&					 	& 							& 	$2_0^14_0^1$	\\
25780			 	& 	2519					&					 	& 							& $2_0^13_0^1$ / $2_0^15_0^2$	\\
26137			 	& 	2876					&					 	& 							& 	$2_0^14_0^2$	\\
26323			 	& 	3062					&					 	& 							& $2_0^13_0^14_0^1$\\\hline
\end{tabular}\\
\label{tab_bands}
\end{table}

\subsection{Gas phase spectrum of HC$_5$H}
\begin{figure}\begin{center}
 \includegraphics[scale=0.4]{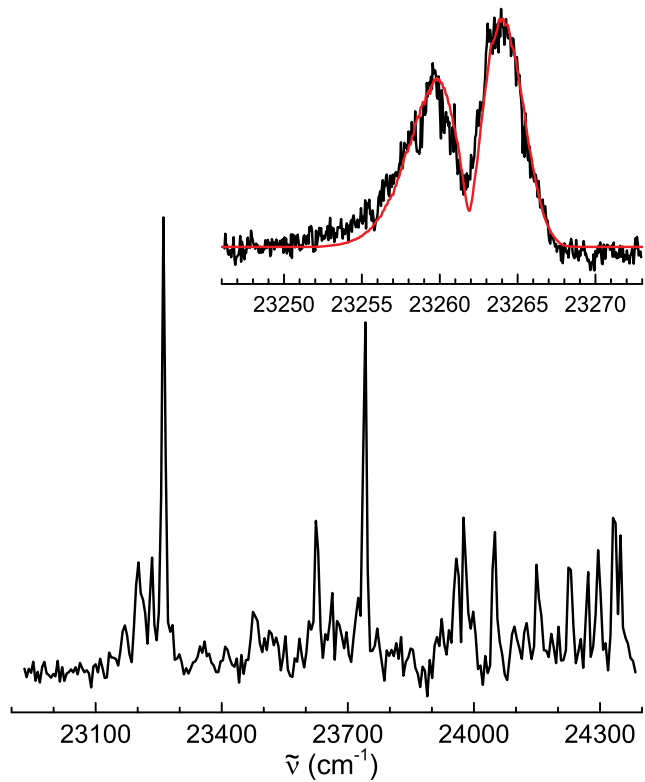}
 \caption{Gas phase $1 ^3\Sigma_u^- \leftarrow X^3\Sigma_g^-$ spectrum of HC$_5$H (0.1\,nm laser bandwidth). Inset: origin band (0.001\,nm resolution; black trace) and calculated rotational profile (T$_{rot} = 40$ K; red).} \label{fig_REMPI}
\end{center}\end{figure}

The $1 ^3\Sigma_u^- \leftarrow X^3\Sigma_g^-$ R2C2PI spectrum measured in the gas phase with 0.1\,nm laser bandwidth is shown in Fig. \ref{fig_REMPI}; the inset is a higher-resolution recording (0.001\,nm) of the origin band at 23261\,cm$^{-1}$ along with a simulated profile at the B3LYP/cc-pVTZ level. The gas phase position of the origin band is identical to the wavenumber of the lower-energy site in the neon matrix. Quite unusually, a blue shift of 5 to 52\,cm$^{-1}$ is observed in neon compared to the gas phase for the other bands and matrix sites. The longer HC$_{2n+1}$H members experience red-shifts in neon that increase with chain length.\cite{fulara95,ball00,ding03} Three hot bands to the red of the origin band, assigned as $4_1^1$, $5_1^1$, and $4_1^1$$5_1^1$, are discernible in the gas phase spectrum. The vibrational temperature is estimated at about 350\,K. In addition to the other bands already observed in neon, four weak, yet unidentified features were found by R2C2PI. Their broadness might mask them in the matrix.

The rotational profile of the origin band is very similar to those of the longer HC$_{2n+1}$H homologues,\cite{ball00} except for its width, which is circa two times broader. It is satisfactorily explained by the calculated rotational constants in the ground and excited state. Theoretical values at the B3LYP and CASPT2 levels are very similar and lie within the error margins of those ones that were determined by a least squares fit (Table \ref{tab_fit}). The rotational temperature is about 40\,K. Spin-spin splitting effects are not discernible. Individual rotational lines are also not resolved. The Lorentzian line width is around 0.5\,cm$^{-1}$, corresponding to an upper electronic state lifetime of 0.1\,ps. This is much shorter than the jitter between both lasers used ($\sim 40$\,ns). Resonance enhanced ionization can only be achieved if after excitation by the scan laser the molecule relaxes within 0.1\,ps to another long-lived state ($\gtrsim 40$\,ns) above the ground one.

\begin{table}
 \caption{Spectroscopic constants (cm$^{-1}$) in the $X^3\Sigma_g^-$ and $1 ^3\Sigma_u^-$ electronic states of HC$_5$H.}
\begin{tabular}{|r|ccc|} \hline
										& \multicolumn{2}{c}{Calculation}& Fit of $0^0_0$		\\
										& B3LYP	& CASPT2				& 						\\ \hline \hline
$\widetilde{v}_{00}$						& 28339	&	22984				& 23261.8$\pm$0.3	\\
B$_{\text{v}}'$						& 0.0761	&	0.0753				& 0.0758$\pm$0.0004	\\
B$_{\text{v}}''$ 						& 0.0754	&	0.0746				& 0.0749$\pm$0.0004	\\ \hline
\end{tabular}\\
\label{tab_fit}
\end{table}

\subsection{Matrix spectrum of H$_2$CCCCC}
\begin{figure}\begin{center}
 \includegraphics[scale=0.4]{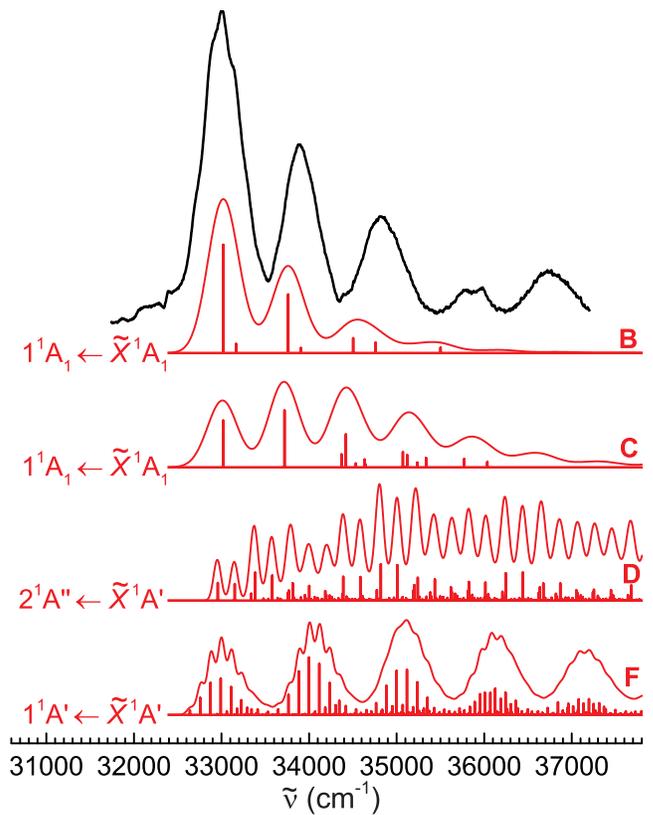}
 \caption{C$_5$H$_2$ absorption spectrum in a neon matrix (black trace) in comparison to Franck-Condon simulation for different isomers at the B3LYP/cc-pVTZ level (T$_{vib}=0$ K; red traces).} \label{fig_UV_matrix}
\end{center}\end{figure}

The other absorption system in a neon matrix commences at 303\,nm (4.1\,eV) and is composed of five broad bands with FWHM of about 450\,cm$^{-1}$ (Fig. \ref{fig_UV_matrix}). It is about twenty times stronger than the 430\,nm HC$_5$H system. Attempts to measure the corresponding R2C2PI spectrum failed. Intense electronic transitions are not expected for the linear triplet around 4.1\,eV (Table \ref{tab_calc}). Another C$_5$H$_2$ isomer is likely the carrier. The first strong transitions of isomers \textbf{B}, \textbf{C}, \textbf{D}, and \textbf{F} are calculated in proximity of the observed absorption energy. Franck-Condon simulations of these transitions were conducted at the B3LYP/cc-pVTZ level. They are compared with the measured spectrum in Fig. \ref{fig_UV_matrix}. In terms of stability, isomer \textbf{D} (= ethynylcyclopropenylidene) and its precursor in the matrix experiment, the \textbf{D}$^+$ cation, are close in energy to \textbf{A} and \textbf{A}$^+$ (see Fig.\,\ref{fig_structures}). However, strong excitation of C--H bending vibrations is predicted to result in a long and congested progression in contradiction to the observations. Furthermore \textbf{D}$^+$ has a calculated transition in the visible with about the same intensity as that of the neutral, which was not observed (Table \ref{tab1SI_calc}). The only cationic absorption system that could be detected in the matrix before neutralization was that of \textbf{A}$^+$. The carrier of the 303\,nm features, if produced via photo-induced neutralization, must have a cationic form with none or weak absorptions in the monitored 260--1100\,nm wavelength range.

The best match between computed Franck-Condon intensities and the observed vibrational progression is provided by the $1 ^1 \textnormal{A}_1 \leftarrow \widetilde{X} ^1 \textnormal{A}_1$ transition of pentatetraenylidene (\textbf{B}). It has an oscillator strength of $f = 0.12$ at the CASPT2 level, while the \textbf{B}$^+$ cation absorbs only weakly ($f < 0.01$) below 6\,eV (Table \ref{tab1SI_calc}). As demonstrated recently, the H$_2$C$_5$H$^+$ cation is efficiently produced from the methyldiacetylene precursor in the same ion source.\cite{fulara15} \textbf{B}$^+$ is easily generated by C--H bond cleavage from H$_2$C$_5$H$^+$. Moreover, \textbf{B} and \textbf{B}$^+$ can also be formed from \textbf{A} and \textbf{A}$^+$ \textit{via} hydrogen migration during the trapping of mass-selected ions. It was recently shown for C$_7$H$_3$ that this is a feasible process.\cite{chakraborty14} Therefore the 303\,nm system is concluded to be caused by the afore-mentioned transition of the cumulene carbene H$_2$C$_5$. Considering the calculated oscillator strengths and observed absorption intensities, \textbf{A} and \textbf{B} had about the same abundances in the matrix.

\begin{table}
 \caption{Calculated electronic transitions of C$_5$H$_2$ isomers below 6\,eV.}
\begin{tabular}{|c|cc|cc|} \hline
& \multicolumn{2}{c|}{SAC-CI}& \multicolumn{2}{c|}{CASPT2 $^a$}		\\ 
State				&	[eV]&\textit{f}			&	[eV]		&\textit{f}		\\\hline \hline
\multicolumn{5}{|c|}{\textbf{A}\hspace{0.3cm} $X^3\Sigma_g^-$}\\ \hline
$1^3\Sigma_u^-$	& 3.93	&\textit{0.063}		& 3.28 (2.85)	&\textit{0.005}	\\\
$2^3\Sigma_u^-$	& 5.46	&\textit{0.002}		& 4.76			&\textit{0.0000}	\\\hline
\multicolumn{5}{|c|}{\textbf{B}\hspace{0.3cm} $\widetilde{X}^1$A$_1$}				\\ \hline
$1^1$A$_1$	 	& 3.73 	&\textit{0.15}		& 4.00 (3.87) 	&\textit{0.12}	\\
$2^1$A$_1$	 	& 5.80	&\textit{0.022}		& 5.02			&\textit{0.0004}\\
$1^1$B$_1$	 	& 2.41	&\textit{0.006}		& 2.03			&\textit{0.004}\\
$2^1$B$_1$	 	& $>$6.0	&					& 4.99			&\textit{0.009}\\
$1^1$B$_2$	 	& 5.78	&\textit{0.031}		& 5.52			&\textit{0.024}	\\\hline
\multicolumn{5}{|c|}{\textbf{C}\hspace{0.3cm} $\widetilde{X}^1$A$_1$}				\\ \hline
$1^1$A$_1$	 	& 4.87	&\textit{0.23}		& 4.65 (4.51)	&\textit{0.22}	\\
$1^1$B$_1$	 	& 3.50	&\textit{0.014}		& 3.21			&\textit{0.012}\\
$1^1$B$_2$	 	& 5.07	&\textit{0.019}		& 4.77			&\textit{0.010}	\\\hline
\multicolumn{5}{|c|}{\textbf{D}\hspace{0.3cm} $\widetilde{X}^1$A$'$}				\\ \hline
$1^1$A$'$	 		& 5.82	&\textit{0.028}		& 5.82 			&\textit{0.020}	\\
$1^1$A$''$	 		& 3.22	&\textit{0.001}		& 3.24			&\textit{0.001}\\
$2^1$A$''$	 		& 4.48	&\textit{0.018}		& 4.66 (4.41)	&\textit{0.025}	\\\hline
\multicolumn{5}{|c|}{\textbf{E}\hspace{0.3cm} $\widetilde{X}^1$A$'$}				\\ \hline
$1^1$A$'$	 		& 3.46	&\textit{0.020}		& 3.58 (2.33)	&\textit{0.024}	\\
$2^1$A$'$	 		& 4.57	&\textit{0.022}		& 4.67			&\textit{0.010}\\
$1^1$A$''$	 		& 2.60	&\textit{0.001}		& 2.44		 	&\textit{0.002}	\\
$2^1$A$''$	 		& 2.87	&\textit{0.003}		& 3.01 			&\textit{0.004}	\\
$3^1$A$''$	 		& 4.90	&\textit{0.0000}	& 5.08			&\textit{0.0003}\\
$4^1$A$''$	 		& $>$6.0	&					& 5.54			&\textit{0.001}	\\\hline
\multicolumn{5}{|c|}{\textbf{F}\hspace{0.3cm} $\widetilde{X}^1$A$'$}				\\ \hline
$1^1$A$'$	 		& 4.27	&\textit{0.36}		& 4.35 (4.23)	&\textit{0.33}	\\
$2^1$A$'$	 		& 6.07	&\textit{0.023}		& 5.69			&\textit{0.025}	\\
$1^1$A$''$	 		& 1.54	&\textit{0.0000}	& 1.76			&\textit{0.0001}\\
$2^1$A$''$	 		& 2.30	&\textit{0.007}		& 2.10			&\textit{0.0047}\\
$3^1$A$''$	 		& 4.15	&\textit{0.0000}	& 4.32			&\textit{0.0001}\\\hline
\end{tabular}\\
\begin{tabular}{l}
$^a$ \footnotesize{vertical energies at MS(4)-CASPT2(10/12);}\\
\textcolor{white}{$^a$} \footnotesize{adiabatic ones in brackets at MS(4)-CASPT2(10/10)}
\end{tabular}
\label{tab_calc}
\end{table}

The proposed band assignments are listed in Table\,\ref{tab2_bands}. The progression is mainly caused by excitations of C--C stretching vibrations $\nu_3$(a$_1$) and $\nu_6$(a$_1$). Band widths are around 200\,cm$^{-1}$, corresponding to the excited electronic state lifetime of about 30\,fs. Calculated vibrational frequencies in excited electronic states are not expected to be as accurate as ground state ones. At the B3LYP/cc-pVTZ level, $\nu_3$(a$_1$) and $\nu_6$(a$_1$) deviate from observed values by $- 17$\% and $- 5$\%, respectively. The Franck-Condon approximation often underestimates the intensities of combination bands. This is also noticed here for the bluest observed band at 36740\,cm$^{-1}$. A faint shoulder is recognizable on the high energy side of the origin band. It is assigned to the double excitation of the in-plane C--H$_2$ bending mode $\nu_{15}$(b$_2$).

\begin{table}
 \caption{Electronic transitions of C$_5$H$_2^+$ isomers below 6\,eV predicted with the MS(6)-CASPT2(11/12)/cc-pVTZ method using ground state geometries from M06-2X/cc-pVTZ calculations.}
\begin{tabular}{|ccc|ccc|ccc|} \hline
\multicolumn{3}{|c|}{\textbf{B}$^+$\hspace{0.2cm} $\widetilde{X}^2$B$_2$} & \multicolumn{3}{|c|}{\textbf{D}$^+$\hspace{0.2cm} $\widetilde{X}^2$A$'$} & \multicolumn{3}{|c|}{\textbf{F}$^+$\hspace{0.2cm} $\widetilde{X}^2$A$'$}\\ \hline
State			&	[eV]&\textit{f}		& State			&	[eV]&\textit{f}		& State				&	[eV]&\textit{f}		\\\hline
$1^2$B$_2$	& 2.59	&\textit{0.001}	& $1^2$A$'$	& 3.01	&\textit{0.029}	&	$1^2$A$'$		& 1.00	&\textit{0.001}	\\
$2^2$B$_2$	& 3.01	&\textit{0.001}	& $2^2$A$'$	& 4.31	&\textit{0.001}	&	$3^2$A$'$		& 3.11	&\textit{0.001}	\\
$3^2$B$_2$	& 3.50	&\textit{0.007}	& $3^2$A$'$	& 5.19	&\textit{0.002}	&					&		&				\\
$4^2$A$_1$	& 4.52	&\textit{0.002}	& $4^2$A$'$	& 5.48	&\textit{0.17}	&					&		&				\\
$1^2$A$_2$	& 1.87	&\textit{0.002}	& $1^2$A$''$	& 3.02	&\textit{0.001}	&					&		&				\\
$2^2$A$_2$	& 2.41	&\textit{0.004}	& $3^2$A$''$	& 4.47	&\textit{0.001}	&					&		&				\\
$3^2$A$_2$	& 4.40	&\textit{0.001}	& $5^2$A$''$	& 5.76	&\textit{0.002}	&					&		&				\\
$4^2$A$_2$	& 4.91	&\textit{0.002}	& 				&		&				&					&		&				\\
$5^2$A$_2$	& 5.25	&\textit{0.006}	& 				&		&				&					&		&				\\\hline
\end{tabular}\\
\label{tab1SI_calc}
\end{table}

\begin{table}
 \caption{Observed absorption maxima (cm$^{-1}$) in the $1 ^1\text{A}_1 \leftarrow \widetilde{X} ^1\text{A}_1$ electronic spectrum of H$_2$CCCCC in a neon matrix and proposed assignment.}
\begin{tabular}{|ccc|} \hline
$\widetilde{\nu}$ 	& $\Delta\widetilde{\nu}$	& Assignment			\\ \hline \hline
33000			 	& 	0						& $0_0^0$		\\
33140			 	& 	140						& $15_0^2$		\\
33890			 	& 	890						& $6_0^1$		\\
34830			 	& 	1830					& $3_0^1$ , $6_0^2$		\\
35860			 	& 	2860					& $3_0^1 6_0^1$ , $6_0^3$	\\
36740			 	& 	3740					& $3_0^2$ , $3_0^1 6_0^2$ , $6_0^4$\\\hline
\end{tabular}\\
\label{tab2_bands}
\end{table}

\section{Summary}
The $1 ^3\Sigma_u^- \leftarrow X^3\Sigma_g^-$ transition of HC$_5$H has been remeasured in a neon matrix and band assignments have been revised. The respective gas phase spectrum has been observed by R2C2PI for the first time. C$_5$ and HC$_5$ have already been found in the circumstellar shell of the carbon star CW Leonis and the Taurus molecular cloud 1.\cite{bernath89,cernicharo86,cernicharo87} A detection of HC$_5$H might be accomplished with the present data, but the theoretical oscillator strength is only 5$\times 10^{-3}$. The $1 ^3\Sigma_u^- \leftarrow X^3\Sigma_g^-$ transition is characterized by a bent excited state structure, resulting in a vibrational progression with strong contribution from a C--H bending mode. The smaller propargylene diradical, HC$_3$H, has recently been found to be quasilinear (C$_2$ symmetry) in its $\widetilde{X}^3$B ground state,\cite{osborn14} with a deformed chain structure which is similarly suggested for HC$_5$H in the $1 ^3\Sigma_u^-$ state.

A UV band system of another C$_5$H$_2$ isomer was also measured in the matrix. It is assigned to the strong $1 ^1 \textnormal{A}_1 \leftarrow \widetilde{X} ^1 \textnormal{A}_1$ transition of pentatetraenylidene H$_2$CCCCC (\textbf{B}) based on calculated excitation energies and Franck-Condon simulations of the most stable isomers.

\begin{acknowledgments}
This work has been funded by the Swiss National Science Foundation (Project 200020-140316/1).
\end{acknowledgments}

\bibliography{references}

\end{document}